%% file: main.tex
\RequirePackage{tikz} 

\documentclass[letterpaper, 10 pt, conference]{ieeeconf}  

\IEEEoverridecommandlockouts                              

\usepackage{comment}        
\usepackage{amsmath,amssymb,amsfonts}
\usepackage{algorithm}
\usepackage[noend]{algpseudocode}
\usepackage{graphicx}
\usepackage{textcomp}
\usepackage{xcolor}
\usepackage{booktabs}
\usepackage{hyperref}
\usepackage{xspace}
\usepackage{balance}
\usepackage{array}      
\usetikzlibrary{arrows,calc,positioning,shapes,fit}

\usepackage{caption}

\tikzstyle{block} = [rectangle, draw, text centered,
  rounded corners, minimum height=2em]
\tikzstyle{eq} = [circle, draw, minimum height=1.7em, inner sep=1pt] 

\makeatletter
\newcommand{\gettikzxy}[3]{%
  \tikz@scan@one@point\pgfutil@firstofone#1\relax
  \edef#2{\the\pgf@x}%
  \edef#3{\the\pgf@y}%
}
\makeatother

\usepackage{biblatex}
\def\BibTeX{{\rm B\kern-.05em{\sc i\kern-.025em b}\kern-.08em
    T\kern-.1667em\lower.7ex\hbox{E}\kern-.125emX}}
\title{\LARGE \bf
Preparation of Papers for IEEE Sponsored Conferences \& Symposia*
}

\title{Prioritized Planning for Continuous-time Lifelong Multi-agent Pathfinding\\
\author{Alvin Combrink$^{1}$, Sabino Francesco Roselli$^{1}$, Martin Fabian$^1$
\thanks{$^{1}$Division of Systems and Control, Department of Electrical Engineering, Chalmers University of Technology, G{\"o}teborg, Sweden
        {\tt\footnotesize \{combrink, rsabino, fabian\}@chalmers.se}}
}
\thanks{We gratefully acknowledge the Vinnova project
CLOUDS (Intelligent algorithms to support Circular soLutions fOr sUstainable proDuction Systems), and the Wallenberg AI, Autonomous Systems and Software Program (WASP) funded by the Knut and Alice Wallenberg Foundation.} 
}

\input{NewCommands}

\begin{document}
\maketitle
\thispagestyle{empty} 
\pagestyle{empty} 
\begin{abstract}
\input{Abstract}
\end{abstract}
\section{INTRODUCTION}
\label{sec:Introduction}
\input{Introduction}

\section{Background}
\label{sec:Background}

\input{Background}

\section{Problem Definition}
\label{sec:ProblemDefinition}

\input{ProblemDefinition}

\section{Method}
\label{sec:Method}
\input{Method}

\section{Experimental Evaluation}
\label{sec:ExperimentalEvaluation}
\input{ExperimentalEvaluation}

\section{Discussion}
\label{sec:Discussions}
\input{Discussions}
\section{Conclusions}
\label{sec:Conclusions}

\input{Conclusions}

\printbibliography

\end{document}

%% file: NewCommands.tex
\newcommand{\GitHubLink}{{\scriptsize\url{https://github.com/Adcombrink/CPLP-LifelongMAPFR}\xspace}}

\newcommand{\LMAPFR}{\ensuremath{\text{LMAPF}_R}\xspace}
\newcommand{\planner}{CPLP\xspace}

\newcommand{\Graph}{\mathcal{G}}
\newcommand{\Vertices}{\mathcal{V}}
\newcommand{\vertex}{v}
\newcommand{\Edges}{\mathcal{E}}
\newcommand{\edge}{e}

\newcommand{\agentspeed}{s}
\newcommand{\agentradius}{r}
\newcommand{\Tasks}{\mathcal{T}}
\newcommand{\task}{\tau}
\newcommand{\plan}{\pi}

\newcommand{\Agents}{\mathcal{A}}
\newcommand{\agent}{a}
\newcommand{\agentStart}{\vertex_s}

\newcommand{\funcAddWait}[1]{\Call{AddWaitActions}{#1}}
\newcommand{\funcTaskCompletion}[1]{\Call{TaskCompletion}{#1}}
\newcommand{\funcTaskPriorities}[1]{\Call{UpdateTaskPriorities}{#1}}
\newcommand{\funcAssignAgents}{\Call{AssignTasksToAgents}{}}

\newcommand{\funcDist}[1]{\Call{ShortestPathTime}{#1}}

\newcommand{\funcPlanHPA}{\Call{PrioritizedPlans}}
\newcommand{\funcPlanShortPaths}{\Call{ShortPlans}}
\newcommand{\funcRandShortPaths}{\Call{ShortRandomPlans}}
\newcommand{\funcNextPlanningTime}{\Call{GetNextPlanningTime}}

\newcommand{\none}{\text{None}\xspace}
\newcommand{\TaskSet}{\Tasks^U}
\newcommand{\newTasks}{\Tasks_{\mathit{new}}}

\newcommand{\HPT}{{\ensuremath{\tau^*}}}
\newcommand{\HPA}{{\ensuremath{\agent^*}}}
\newcommand{\currTime}{t_{\mathit{plan}}}
\newcommand{\nextTime}{t_{\mathit{next}}}
\newcommand{\marginTime}{\Delta t}

\newcommand{\horizon}{\bar{t}}

\newcommand{\prioritizedAgentLimit}{\alpha}
\newcommand{\numAgents}{n_\agent}
\newcommand{\numVertices}{n_\vertex}
\newcommand{\vertexPerAgent}{\rho}
\newcommand{\paramAdditionalVertices}{\gamma_1}
\newcommand{\releaseRate}{r}

%% file: Abstract.tex
Multi-agent Path Finding (MAPF) is the problem of planning collision-free movements of agents so that they get from where they are to where they need to be. Commonly, agents are located on a graph and can traverse edges. This problem has many variations and has been studied for decades. Two such variations are the \emph{continuous-time} and the \emph{lifelong} MAPF problems. 
In the former, edges have non-unit lengths and volumetric agents can traverse them at any real-valued time.
In the latter, agents must attend to a continuous stream of incoming tasks.
Much work has been devoted to designing solution methods within these two areas. 
To our knowledge, however, the combined problem of continuous-time lifelong MAPF has yet to be addressed.

This work addresses continuous-time lifelong MAPF with volumetric agents by presenting the fast and sub-optimal Continuous-time Prioritized Lifelong Planner (\planner).
\planner continuously assigns agents to tasks and computes plans using a combination of two path planners; one based on CCBS and the other based on SIPP.
Experimental results with up to $800$ agents on graphs with up to $12\,000$ vertices demonstrate practical performance, where maximum planning times fall within the available time budget.
Additionally, \planner ensures collision-free movement even when failing to meet this budget. Therefore, the robustness of \planner highlights its potential for real-world applications.



%% file: Introduction.tex
Multi-agent Path Finding (MAPF) is the problem of planning collision-free movements of agents to get them from where they are to where they need to be. 
Many methods exist for the many variants of the problem. 
In its most common form, time is discretized and agents are represented by points moving on a graph. Each edge takes unit-time to traverse and all agents move in lockstep. If any two agents occupy the same vertex or traverse the same (undirected) edge in opposite directions at the same time, a collision occurs~\cite{ SurveyMa2022,SurveyStern2019}. 
This problem has been studied for decades~\cite{erdmann1987} and is NP-hard to solve for minimum \emph{makespan}, \emph{sum-of-arrival-times} and travelled distance~\cite{NPhardness}. 

Optimal solvers include M$^*$~\cite{wagner2011m}, Increasing-cost tree search~\cite{ICTS}, and the seminal Conflict-based Search (CBS)~\cite{CBS}. 
Among the sub-optimal, heuristic-based solvers are 
Windowed Hierarchical Cooperative A*~\cite{CoopPF},
bounded sub-optimal CBS~\cite{barer2014suboptimal} and 
Priority Inheritance with Backtracking (PIBT) for fast planning of large numbers of agents with certain theoretical guarantees~\cite{PIBT}.

In recent years, much focus has been directed toward the continuous-time variant MAPF$_{R}$, where edges can take any positive time to traverse and agents can traverse them at any time. In many cases, volumetric agents are considered.
Safe Interval Path Planning (SIPP)~\cite{SIPP} underlies many solution methods for MAPF$_R$, such as Continuous-time CBS (CCBS)~\cite{CCBS}, Satisfiability Modulo Theory-CBS~\cite{surynek2019multi} and PSIPP~\cite{PSIPP}, amongst others.

Another variant of MAPF is the \emph{lifelong}, or \emph{online}, version. Here, agents must attend to a stream of incoming and a priori unknown tasks, which are completed when an agent occupies a respective task's target vertex some time after the task has been released to the system. 
Agents are not assumed to always be assigned a task, thus, methods for this problem must address idle agents. Substantial work has been done in this area too, such as token passing~\cite{tokenpassing}, Rolling horizon collision avoidance~\cite{RHCR}, FM-scheduler~\cite{FMScheduler},
Primal$_2$ which uses reinforcement learning for decentralised path planning in partially observable environments~\cite{damani2021primal},
and~\cite{concurrentplanning} for concurrent planning and executing in an online setting, where plans are refined for as long as time constraints allow.


Most industrial automation systems (warehouses, assembly lines, etc.) run continuously and indefinitely, in spaces that cannot be captured by graphs with unit-length edges. The lifelong and continuous-time MAPF variants each address one aspect of this, however, the combined problem of lifelong MAPF in continuous-time (LMAPF$_R$) remains largely unexplored. 
To fill this gap, we introduce the \emph{Continuous-time Prioritized Lifelong Planner} (\planner), a fast, sub-optimal solver for the collision-free planning of volumetric agents on a graph in metric 2D space for lifelong MAPF.


The outline of the article is as follows: relevant background is given in Section~\ref{sec:Background}, followed by the problem definition in Section~\ref{sec:ProblemDefinition}. The planner is described in Section~\ref{sec:Method} and experimentally evaluated in Section~\ref{sec:ExperimentalEvaluation}. Finally, Section~\ref{sec:Discussions} includes discussions and Section~\ref{sec:Conclusions} concludes the article.

%% file: Background.tex
Here we introduce SIPP~\cite{SIPP}, CCBS~\cite{CCBS}, and graph pre-computations~\cite{PSIPP, walkersturtevant} that the following work builds on.

\subsection{SIPP}
\label{sec:background:SIPP}

Safe Interval Path Planning (SIPP)~\cite{SIPP} is a method to plan in continuous time the motion of an agent in an environment with dynamic obstacles. 
If a dynamic obstacle occupies a vertex or edge during a certain time interval, 
then it is unsafe for the agent to also do so at any time within this interval as that would cause a collision.
Thus, the combined movement of all obstacles provides a sequence of alternating safe and unsafe time intervals for each vertex and edge.
A collision-free trajectory from an agent's current location and time to a target location can be planned by performing an A$^*$~\cite{A_star} search in the vertex-safe interval space.  

\subsection{CCBS}

Continous-time Conflict Based Search (CCBS)~\cite{CCBS} extends CBS~\cite{CBS} to the MAPF$_R$ problem. 
Given a set of agents on a graph, each with their respective start and target vertices, CCBS finds an optimal collision-free trajectory for each agent. 
CCBS's high-level algorithm is a best-first search in a binary constraint-tree. 
A node $N$ in the tree represents a set $N_\Pi$ containing one trajectory for each agent, and a set of constraints $N_c$.
At the root node $N^r$, $N^r_\Pi$ contains for each agent the shortest path from its start vertex to target vertex without considering other agents, and $N^r_c = \varnothing$. $N^r$ is inserted into the open set.
At each iteration of the high-level search, a node $N$ minimizing an objective function over all nodes in the open set is selected.
If no collisions are detected between the trajectories in $N_\Pi$, then $N_\Pi$ is returned as the solution. 
However, if a collision is detected, say between agent $a_1$ performing an action $m_1$ at time $t_1$ and $a_2$ performing action $m_2$ at time $t_2$, then two new nodes $N^1$ and $N^2$ are spawned. 
For $i=1,2$, $N^i_c = N_c \cup \{c_i\}$ where $c_i$ forbids $a_i$ from performing $m_i$ within the interval $[t_i, t^u_i)$, where $t^u_i$ is the earliest time where $m_i$ can be performed without colliding with the other agent performing its action.
All trajectories in $N_\Pi$ are copied into $N^i_\Pi$ except for $a_i$'s path which is recomputed using CSIPP (the variant of SIPP used in \cite{CCBS}) to satisfy $N^i_c$.


\newcommand{\EVEC}{\ensuremath{\Edges_\mathit{VEC}}\xspace}
\newcommand{\VVEC}{\ensuremath{\Vertices_\mathit{VEC}}\xspace}
\newcommand{\EEEC}{\ensuremath{\Edges_\mathit{EEC}}\xspace}

\subsection{Graph Pre-computation}
Determining if arbitrary geometric shapes overlap can be computationally expensive, which is particularly undesirable in online systems. However, by assuming that all agent volumes are described by circles with the same radius and move with the same speed,~\cite{PSIPP, walkersturtevant} describe how unsafe intervals for edge-vertex and edge-edge pairs can be pre-computed. 
For practical roadmaps, computations can be done in near $\mathcal{O}\left(\left| \Vertices \right| + \left| \Edges \right|\log \left| \Edges \right| \right)$ and then stored in lookup tables for use during runtime.

%% file: ProblemDefinition.tex
An \LMAPFR problem is a tuple $\langle \Graph, \Agents, \Tasks, \agentStart \rangle$ containing a graph $\Graph = \langle \Vertices, \Edges \rangle$, agents $\Agents$, tasks $\Tasks$, and a mapping $\agentStart: \Agents \rightarrow \Vertices$ defining the agents' respective start vertices. 

$\Graph$ is connected and directed, with $\Vertices\subseteq\Re^2$ being points in 2D space and edges $\Edges \subseteq \Vertices\times\Vertices$ connecting vertices.
Agents move along edges with the same constant speed $\agentspeed$ and have circular shapes with radius $\agentradius$. 
When traversing an edge $\edge = \langle \vertex_1, \vertex_2 \rangle \in \Edges$, agents travel in a straight line from $\vertex_1$ to $\vertex_2$, taking 
$\frac{|\vertex_1 - \vertex_2|}{s}$
time to do so.
A collision occurs when two agents' volumes overlap. In this case of circular agents, that is when the distance between the positions of two agents is less than $2\,\agentradius$.

$\Tasks$ is defined as a, possibly unbounded, multiset over $\Vertices\times\Re$, allowing for multiple instances of the same task. A task $\task = \langle \vertex_\task, t_\task \rangle \in \Tasks$ specifies a target vertex $\vertex_\task$ and a release time $t_\task$. For $\task$ to be completed, an agent must be located at $\vertex_\task$ at some time $t \geq t_\task$. Knowledge of $\task$ is only released to the system at $t = t_\task$.

We define a \emph{move-action} as a tuple $\langle \edge, t \rangle \in \Edges \times \Re$ (starting to traverse an edge $\edge$ at time $t$) and a \emph{wait-action}  as a triple $\langle \vertex, t_1, t_2 \rangle \in \Vertices\times\Re^2$ (occupying vertex $\vertex$ from time $t_1$ until $t_2$). 
A plan $\plan_\agent$ for an agent $\agent$ is a sequence of move and wait-actions.

A satisfying solution to an LMAPF$_R$ problem provides a collision-free plan for each agent $\agent\in\Agents$, starting at vertex $\agentStart(\agent)$, such that all tasks in $\Tasks$ are completed as time $t\rightarrow\infty$.
Throughput is defined as the number of completed tasks per time unit, and is typically used as a measurement of solution quality. In real-world settings, however, computation time is also a critical measurement to consider.

%% file: Method.tex
In this section, we present the \emph{Continuous-time Prioritized Lifelong Planner} (\planner) and the two path-planners\footnote{The source code, experimental setup, animations, and other supplementary details can be found at \GitHubLink\label{footnote:github}.}.

\subsection{Continuous-time Prioritized Lifelong Planner}

Agent plans are initially empty, $\forall\agent\in\Agents: \plan_\agent = \langle\;\rangle$, and extensions to these plans are computed by repeatedly calling \planner as tasks enter the system. 
When called, \planner computes plans starting at $\marginTime$ in the future which are then appended to the end of the agent plans. 
Thus, actions are never removed.
A sufficiently large choice of $\marginTime$ provides enough time for \planner to compute plans before they start.

Concretely, at time $t$, $\text{\planner}(\currTime, \newTasks)$ is called with $\currTime=t+\marginTime$ and $\newTasks=\left\{\task\in\Tasks \mid t' < t_\task \leq t \right\}$ where $t'$ is the last time the planner was called.
A set of uncompleted tasks is updated, $\TaskSet\gets\TaskSet\cup\newTasks$, and each task is assigned a priority.
Based on these priorities, a specific task $\HPT = \langle \vertex_\HPT, t_\HPT \rangle$ and an assigned agent $\HPA$ is prioritized by computing an extension to $\HPA$'s plan to complete $\HPT$ while moving all idle agents out of the way.
The remaining agents are respectively assigned a task, and a short plan within a time horizon $\horizon$ is quickly computed to move them \emph{toward} the task vertex without necessarily reaching it. 
If the computed plans complete all tasks in $\TaskSet$, then \none is returned; \planner is called next time a new task enters the system.
Otherwise, \planner returns $\nextTime$ from which time new plans are required; \planner is called at $\nextTime - \marginTime$ (such that $\currTime=\nextTime$). 

Unlike PSIPP~\cite{PSIPP} which computes an entire plan for each agent in order of priority, \planner computes an entire plan for only one agent and the remaining agents get a short plan. This has two advantages in this lifelong setting: First, plans based on old information are avoided. New tasks entering the system can be acted upon earlier when decided plans are shorter. 
Second, by computing shorter plans in less time more often, the computation is spread out over more calls to the planner. This can offer more predictability in the computation time, which is generally advantageous in online settings.

We regard \planner as more similar to PIBT~\cite{PIBT}, despite PIBT working with discrete time while PSIPP works with continuous time. 
In PIBT, one-step movements are planned for all agents at every time-step, in order of agent priority.
The highest priority agent remains as such until it reaches its target vertex. 
Additionally, with minor assumptions on the graph, the algorithm guarantees that the highest priority agent will at every time-step move along its preferred edge. By recursion, the highest priority agent will eventually reach its target vertex.
Since the highest priority agent is always an agent that has yet to reach its target vertex, reachability (defined by~\cite{PIBT} as each agent reaching its assigned target vertex within a bounded time limit) is guaranteed.
However, it is not obvious how to directly apply PIBT to the continuous-time setting with agent volumes,
as the core algorithm relies on (1) all agents moving in lockstep, and (2) an agent's movement disrupts at most one other agent. In \LMAPFR, agents do not generally move in lockstep since edges have non-unit lengths, and an agent's movements can potentially disrupt several other agents due to their volumes. 


\planner employs two primary strategies.
First, we assume that all agents remain idle indefinitely once reaching the end of their plan. 
The implication of this is that a plan for agent $\agent$ cannot be decided if it intersects the final position of some other agent $\agent'$, unless a plan to move $\agent'$ out of the way is simultaneously decided.
We cannot guarantee that such a plan exists for $\agent'$.
Thus, if $\agent$'s plan is decided without also verifying that a plan for $\agent'$ to avoid a collision exists, then collision-free movement is not guaranteed.
Therefore, through this strategy we ensure that no agents collide so long as they follow their decided plan and then remain idle.

Second, much like PIBT, one task-agent pair is prioritized over all others. 
However, unlike PIBT which can rely on graph assumptions to ensure that one-step movements will take the prioritized agent to its task vertex, \planner does not.
Instead, \planner computes the entire plan for the prioritized agent all the way to its task vertex, while moving idle agents out of the way.
Using a solution-complete planner ensures that such a set of plans will be found if it exists.
How CCBS is applied for this is discussed in Section~\ref{sec:method:CCBS_planner}, however, we do not guarantee solution-completeness of our implementation.
Given that all agents could in the worst case come to a rest at their last planned positions without collision, if the graph is well-connected and there is sufficient space for agents to maneuver (considering their volumes and the graph's geometry), then we postulate that such a set of plans exists.

\input{Algorithm}

The Pseudo-code for \planner is presented in Algorithm~\ref{alg:high-level}. 
On lines~\ref{alg:CPLP:1_1}-\ref{alg:CPLP:updatePriorities}, variables are updated with $\currTime$ and $\newTasks$: 
tasks in $\newTasks$ are added to $\TaskSet$;
\funcTaskCompletion{} removes all tasks from $\TaskSet$ that will be completed by existing agent plans;
if there are no remaining uncompleted tasks in $\TaskSet$ then $\none$ is returned since there is no need to plan;
\funcAddWait{} extends every agent's plan that ended before $\currTime$ with a wait-action at its last vertex until $\currTime$;
and \funcTaskPriorities{} updates the priorities of the tasks. Any prioritization scheme can be used in \funcTaskPriorities{}, however, care must be taken to avoid task starvation. A simple scheme is used here; 
a task's priority is proportional to the time it has gone without completion.

Only one task $\HPT$ is prioritized at any given time. Thus, $\HPT\gets\none$ at the time when $\HPA$ arrives at $\vertex_{\HPT}$. 
If no task is prioritized at time $\currTime$ (line~\ref{alg:CPLP:HPA_check}), \funcPlanHPA{} on line~\ref{alg:CPLP:CCBS} 
selects a new prioritized task-agent pair and computes plans for $\HPA$ to complete $\HPT$ and all idle agents to move out of the way: 
For each task in $\TaskSet$ (by descending priority), the agent with the earliest arrival time at the task vertex (when traversing the shortest path after following its existing plan) is selected.
For this task-agent pair, the path planner in \funcPlanHPA{} (Section~\ref{sec:method:CCBS_planner}) is invoked to find a valid set of plans.
The path planner operates under a time limit to ensure that the computational budget in this online setting is not exceeded.
If the path planner successfully finds a set of plans within the time limit, then the task-agent pair is prioritized and the plans are used; otherwise, the next task is considered. 
If no plans are found for any task, the process repeats with agents ranked by subsequent earliest arrival times, until the agent with the $\prioritizedAgentLimit^\text{th}$ earliest arrival time. We leave $\prioritizedAgentLimit$ as an algorithm parameters, where a small $\prioritizedAgentLimit$ reduces the chance to find a valid set of plans but also limits the worst-case computation time. 
Within the example set of our experiments,
when no solution was found (line~\ref{alg:CPLP:NoSolution}), it was sufficient to generate short random paths for all agents with \funcRandShortPaths{} (line~\ref{alg:CPLP:RandomPlans}); this might not be generally true, though.


On line~\ref{alg:CPLP:TaskAssignment}, \funcAssignAgents{} orders tasks in descending priority and assigns to each task the agent with the earliest possible arrival time at the task vertex, when following its existing plan and then traversing the shortest path. An agent can only be assigned one task.

On line~\ref{alg:CPLP:SIPP}, \funcPlanShortPaths{} uses a SIPP-based path planner (Section~\ref{sec:method:SIPP_planner}) to compute a short plan for each agent \emph{toward} its respective assigned task vertex. Agents are ordered in descending priority of their respectively assigned task. For each agent, the SIPP-based path planner is called to find an extension to the agent's existing plan that takes the agent closer to its task vertex. If the agent reaches its task, then a new unassigned task is assigned to the agent. This is repeated until either the path planner is unable to find a plan or the agent's plan extends beyond the planning horizon (i.e. the plan ends at some time $t\geq \currTime + \horizon$).

The next time \planner should be called, $\nextTime$, is determined in \funcNextPlanningTime{} and returned on line~\ref{alg:CPLP:return}. 
If all tasks in $\TaskSet$ are scheduled for completion, the maximum end time across all agent plans is returned, as no further planning is needed. This ensures any remaining tasks in $\TaskSet$ can be removed at that time. 
If not all tasks are scheduled for completion, further planning is required. 
In this case, the minimum end time of all agent plans is considered, as it is at that time when an agent becomes available for further planning. 
Since some agents' plans may not have been extended beyond the horizon $\horizon$ due to no valid plans being found, calling \planner again is unlikely to yield solutions for these agents. Therefore, the returned value is the minimum end time of plans ending at $t\geq\currTime+\horizon$. 
If no such plan exists, $\currTime+\horizon$ is returned instead.

\subsection{CCBS-based Path Planner}
\label{sec:method:CCBS_planner}

Unlike in CCBS, which is designed for the \emph{offline} MAPF$_R$ problem,
\planner plans only one agent
to its target vertex while all other agents are idle. 
The CCBS algorithm does not natively handle idle agents, that is, agents without an assigned target vertex. 
In \funcPlanHPA{} on line~\ref{alg:CPLP:CCBS} of Algorithm~\ref{alg:high-level}, a task-agent pair $\langle \task, \agent \rangle$ is selected. For this pair, the CCBS-based path planner is called where the root of the high-level CCBS search is initialized with the shortest path from $\agent$'s last location to $\vertex_{\task}$. For all other agents, it assigns an infinite wait-action $\langle \vertex', t', \infty \rangle$, where $\vertex'$ is the agent's last location and $t'$ the last planned time there.
If a collision occurs between a moving agent and an idle agent $\agent_i$, we compute a new path for $\agent_i$ using CSIPP. However, instead of finding a path to a target vertex (as in the original CSIPP), we search for a path with the earliest departure time away from the idle agent's current vertex to another vertex where no other agent is scheduled to arrive at after. 
The constraint-tree node where the collision was detected is reinserted into the open set with the updated path for $\agent_i$ and no additional constraints.

\subsection{SIPP-based Path Planner}
\label{sec:method:SIPP_planner}

The SIPP implementation, with a few modifications, follows that of~\cite{SIPP, PSIPP}. 
Recall that this planner is used to find a short plan for agent $\agent$ \emph{toward}, but not necessarily \emph{to}, its target vertex $\vertex_\task$. 
Thus, in the A$^*$ search of SIPP, where states are 
pairs of vertices and safe intervals,
the value of a state $\langle \vertex, [t_1, t_2] \rangle$ is equal to $t + \funcDist{\vertex, \vertex_{\task_\agent}}$ where $t\in [t_1, t_2]$ is the arrival time at $\vertex$ and $\funcDist{\vertex, \vertex_{\task_\agent}}$ is the shortest time to traverse from $\vertex$ to $\vertex_{\task}$. 
If $\agent$ is not assigned a task ($\task = \none$), then the value of the state is simply $t$.
For a state $\langle \vertex, [t_1, t_2] \rangle$ to be a goal state, it cannot be the root state and $t_2 = \infty$ such that the agent can remain there indefinitely.

Additionally, recall that no agent is planned to intersect with another agent's last planned position at any time after it arrives there. This is not considered in the CCBS-based path planner since such collisions are inherently handled by adding constraints.
In this path planner, 
however, this is handled when collecting successor states in the SIPP search; any edge and successor that intersects with an agent's last planned position is removed.

%% file: Algorithm.tex
\begin{algorithm}
\caption{Continuous-time Prioritized Lifelong Planner}
\begin{algorithmic}[1]
    
\Procedure{\planner}{$\currTime, \newTasks$}
    \label{alg:high-level}
    \State
    
    \State $\TaskSet \gets \TaskSet \cup \newTasks$ \label{alg:CPLP:1_1}
    \State \funcTaskCompletion{}
    \If{$\TaskSet = \varnothing$}
        \State \Return $\none$                      \label{alg:CPLP:1_2}
    \EndIf
    \State \funcAddWait{}                           \label{alg:CPLP:addWait}
    \State \funcTaskPriorities{}                    \label{alg:CPLP:updatePriorities}
    \State

    \If{$\HPT = \none$}                                         \label{alg:CPLP:HPA_check}
        \State \funcPlanHPA{}           \label{alg:CPLP:CCBS}
    \EndIf
    \State
    \State \funcAssignAgents{}                  \label{alg:CPLP:TaskAssignment} 
    \If{$\HPT = \none$}  \label{alg:CPLP:NoSolution}                       
        \State \funcRandShortPaths{}            \label{alg:CPLP:RandomPlans}
    \Else
        \State \funcPlanShortPaths{}                \label{alg:CPLP:SIPP}
    \EndIf
    \State
    \State \Return \funcNextPlanningTime{}      \label{alg:CPLP:return}
\EndProcedure
\end{algorithmic}
\end{algorithm}

%% file: ExperimentalEvaluation.tex
All experiments are run in Python 3.11 on a 2020 MacBook Air, Apple M1, 16 GB RAM, macOS Sequoia 15.3\footref{footnote:github}.

\subsection{Instance Generation}

\begin{figure}
    \centering
    \includegraphics[width=0.7\linewidth]{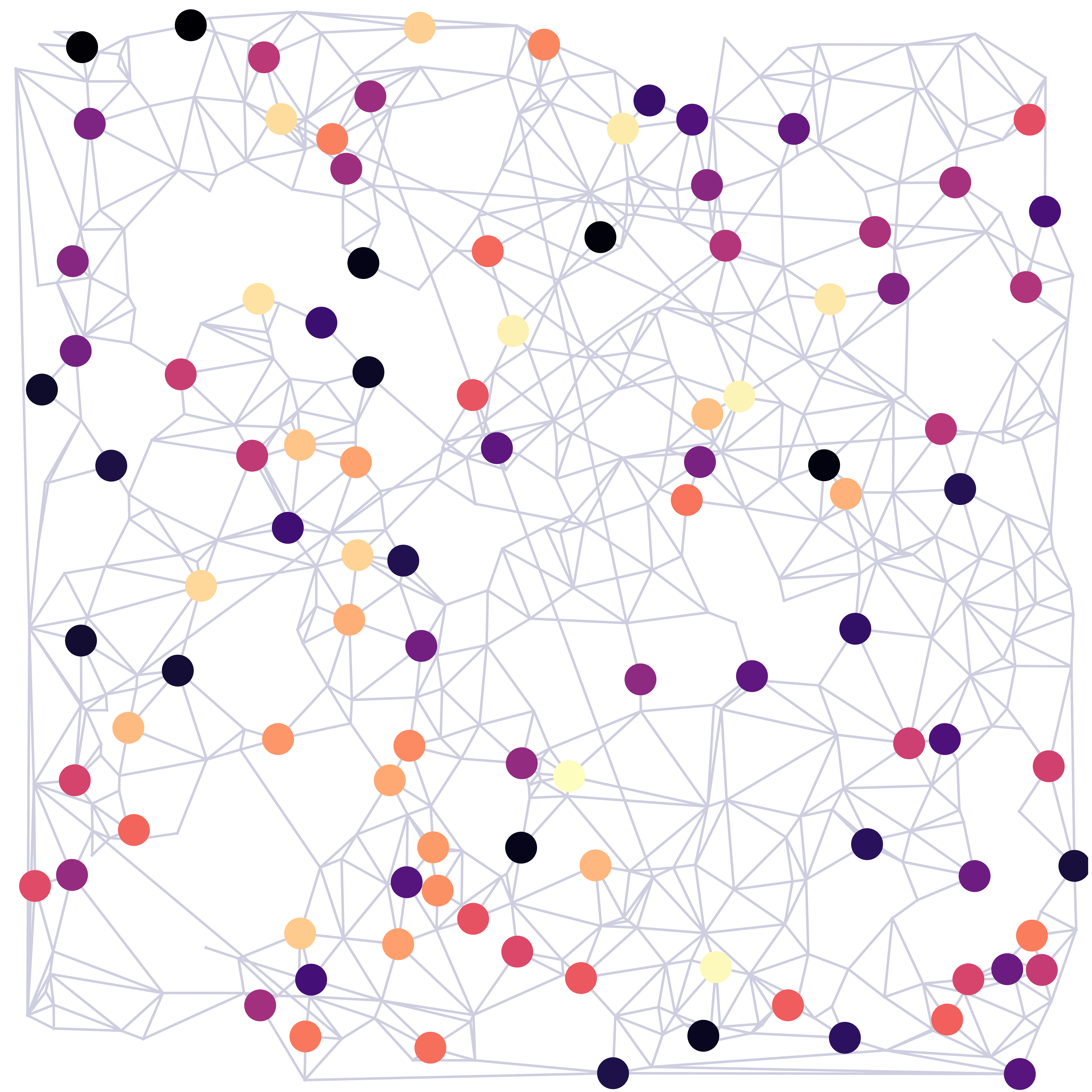}
    \caption{A generated graph with $100$ agents and $500$ vertices. See our repository for animations of similar examples\footref{footnote:github}.}
    \label{fig:RandomInstance}
\end{figure}

Problem instances are generated from a given number of agents $\numAgents$ and vertices per agent $\vertexPerAgent$, so that the resulting graph has $\numVertices = \numAgents \vertexPerAgent$ vertices.
To do so, $\lceil (1+\paramAdditionalVertices) \numVertices \rceil$ 2D points are uniformly sampled from an $l$-by-$l$ space, forming a preliminary set of vertices. Setting $l=3\sqrt{\numVertices}$ results in all generated graphs having the same vertex density.
The vertices' Voronoi cells are then used to create edges between them; any two vertices with bordering cells are connected.
To avoid overly uniform graphs, $\lceil\gamma_1 n_\vertex\rceil$ sampled vertices are removed and the graph is ensured to remain connected.
Finally, $\lceil \gamma_2 n_\vertex \rceil$ sampled vertex pairs are connected to create crossing edges.
In all experiments, undirected edges are used.
The parameters $\gamma_1$ and $\gamma_2$ are set to $0.2$ and $0.02$, respectively.
Agent starting positions are sampled from the set of vertices, ensuring that no two agents start in a collision. Agent radius and speed are both set to $1$. Fig.~\ref{fig:RandomInstance} shows a generated graph with agent starting positions.

Although the lifelong MAPF conceptually never ends, for practical reasons, a finite task set is generated with each task being released within a finite time window $[0, T]$. 
Tasks are released at a rate of $\releaseRate\numAgents$, where $\releaseRate$ is the release rate per agent.
The total number of tasks released is then $\lceil \releaseRate\numAgents T \rceil$.
We select $\releaseRate=0.05$ so that a task is released on average every $20$ seconds per agent, and $T=200$. 
Each respective task $\langle \vertex, t \rangle$ is generated by uniformly sampling a vertex $\vertex\in\Vertices$ and time $t\in\left[0, T\right]$. 
Infeasible instances are manually removed, details are provided in the supplementary materials\footref{footnote:github}.

\subsection{Results}

We ran $15$ instances for each $\numAgents\in[10, 25, 50, 100,$ $200,\dots, 800]$ and $\vertexPerAgent\in[5, 10, 15]$.
Since \planner must compute more plans as $\numAgents$ grows, requiring more computation time, we find $\marginTime = \max\left(\numAgents^{1.25}, 500\right)$~ms to be sufficient.
For \funcPlanHPA{}, we set $\prioritizedAgentLimit=5$ and a time limit of $25$~ms. 
For \funcPlanShortPaths{}, we set horizon $\horizon=1$~s.

\begin{figure}
    \centering
    \includegraphics[width=1\linewidth]{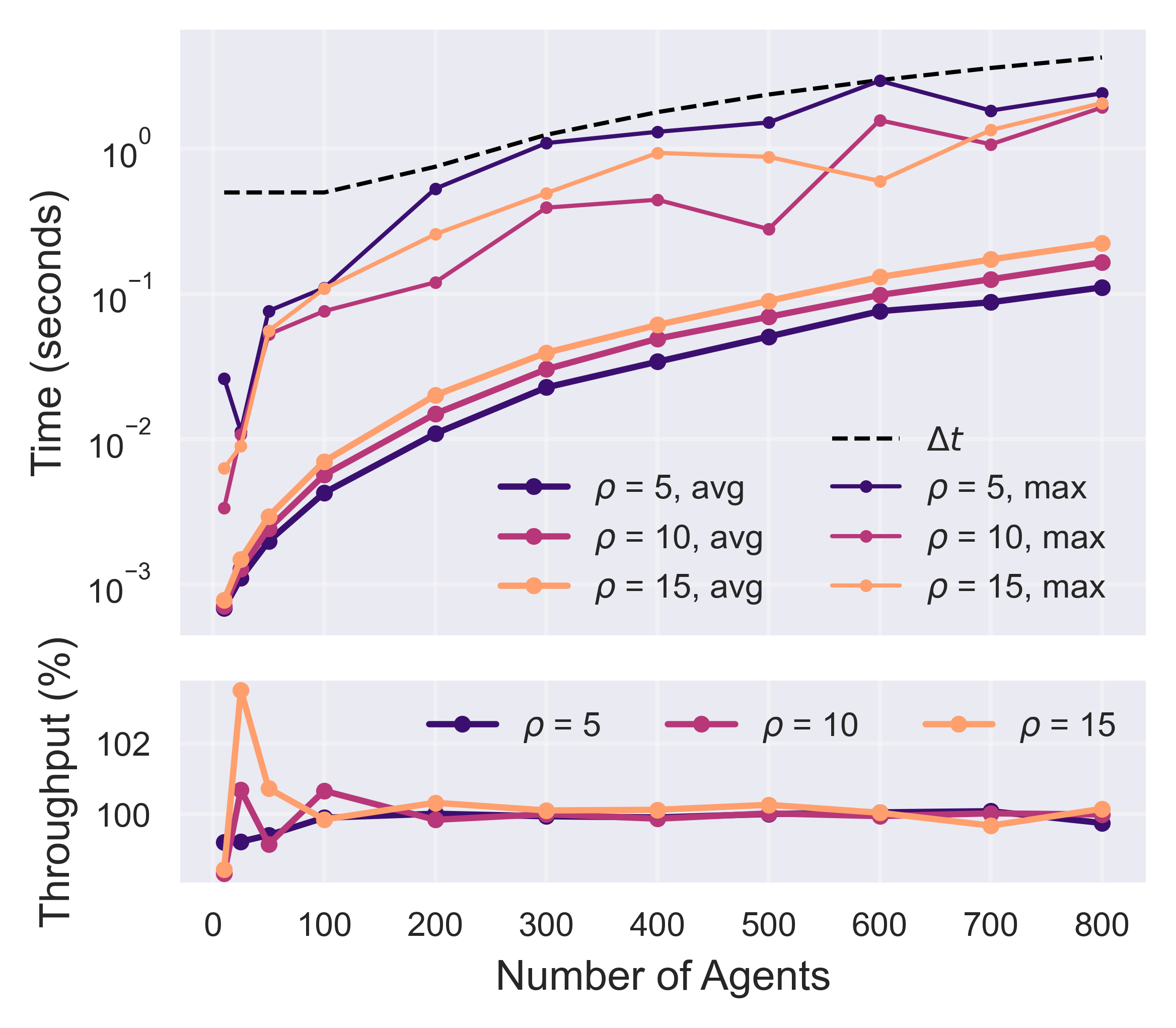}
    \caption{Average and maximum computation time per call to \planner and the average percentage of tasks completed versus released during the time interval $[100, 200]$, for each number of agents and vertices per agent $\vertexPerAgent$.}
    \label{fig:Computation_per_call}
\end{figure}

Fig.~\ref{fig:Computation_per_call} shows the average and maximum computation time per \planner call, and the average percentage of tasks completed versus released during the time interval $[100, 200]$.
Importantly, we see that no call to \planner took more than $\marginTime$. 
Thus, with an appropriate $\marginTime$, \planner can be used in practice without exceeding its computational time budget.
The average computation time remains below $250$~ms for up to $800$ agents, which compared to $\horizon$ shows that plans can be computed in shorter time than their durations. In practice, this means that non-stop movement of agents toward tasks can be maintained.
The throughput exhibits some randomness due to the variability of task release times, particularly when $\numAgents$ is low and fewer tasks are released overall. However, the percentage of completed tasks remains near $100$\%, suggesting that a higher throughput than the tested value ($\releaseRate=0.05$) can be maintained.  

\begin{figure}[]
    \centering
    \includegraphics[width=1\linewidth]{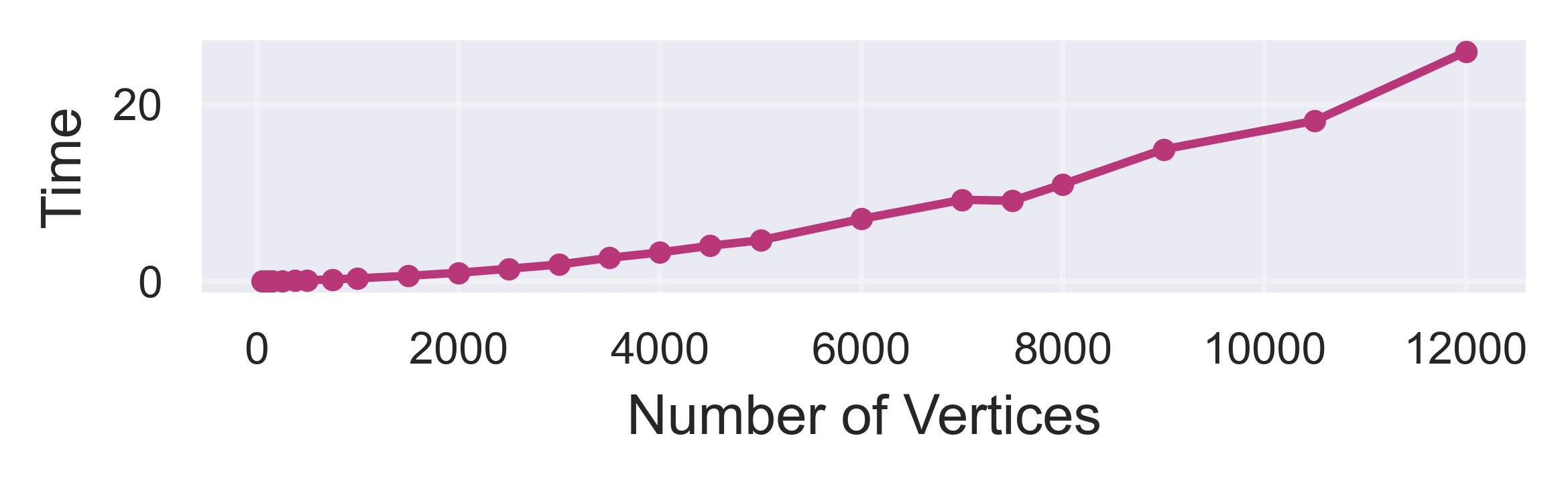}
    \caption{The average graph pre-computation time, in minutes.}
    \label{fig:GPP}
\end{figure}

Fig.~\ref{fig:GPP} shows the average time to perform graph pre-computations, which theoretically grows near $\mathcal{O}\left(\left| \Vertices \right| + \left| \Edges \right|\log \left| \Edges \right| \right)$, although appears near linear at these numbers of vertices.
For the largest graphs tested, containing $12\,000$ vertices, pre-computations took around $35$ minutes on average. These times are manageable in practical settings where agents move on a single, constant graph that only requires pre-computing once.

%% file: Discussions.tex
The proposed \planner demonstrates computation times within practical ranges for hundreds of agents on graphs with thousands of vertices. 
Additionally, the parameters $\horizon$ and $\marginTime$ allow for tuning the planner based on observed average and maximum computation times.
In a real-world setting, $\marginTime$ plays an important roll as it determines the time in advance that \planner must be called, so that the planned paths are computed before they begin.
Increasing $\marginTime$ delays the system's response to new tasks but provides more time to plan.
In our experiments, a $\marginTime$ under $5$~s for the case of $800$ agents is likely manageable in most real-world applications.
An avenue for future work could be to explore adjusting both $\marginTime$ and $\horizon$ adaptively to observed computation times.

Although rarely, \funcRandShortPaths{} was called in a number of instances when a path for $\HPA$ could not be found by \funcPlanHPA{} within the time limit.
Shuffling agents randomly was experimentally sufficient for \funcPlanHPA{} to eventually find a path, however, this might not always work.
In fact, \funcRandShortPaths{} was called indefinitely on infeasible instances where no valid paths exist\footref{footnote:github}.
Additionally, such inefficient behavior is likely not desirable in practice, nor necessary to shuffle \emph{all} agents.
This opens up multiple avenues for future research. 
For instance, the time limit on \funcPlanHPA{} could be temporarily raised, or only agents within the vicinity of the prioritized task-agent pair could be randomly moved instead of moving all agents.
Providing theoretical guarantees for \planner, similar to PIBT, likely hinges on the path planner in \funcPlanHPA{}. 
CCBS was used for this, however, recent findings~\cite{CCBS_revisit} highlight important considerations regarding CCBS's terminatability and optimality trade-offs.
Thus, the exploration of alternative path planners is motivated.


Finally, the problem of lifelong MAPF 
as defined here may have limited practical use. The most common warehouse logistics problem would seem to concern transporting goods from one location to another, that is, \emph{pickup-and-delivery}~\cite{tokenpassing}.
Lifelong MAPF planners can be adapted 
for pickup-and-delivery
by forcing task assignments (as done in~\cite{RHCR}), however,
considering \emph{all} task locations could result in better solutions. Future work could look into adapting \planner for these types of problems, where, for instance, the prioritized agent's entire trajectory through all task locations could be planned while still only computing short plans for the remaining agents. 

%% file: Conclusions.tex
This work presented \planner, a fast, sub-optimal solver for the continuous-time lifelong MAPF problem with agent volumes. \planner combines CCBS and SIPP-based path planning to compute agent plans online, while ensuring collision-free movement even in cases where \planner is unable to compute plans within time constraints. 
Experimental results demonstrate computation times within practical ranges for up to $800$ agents on graphs with up to $12\,000$ vertices. These findings highlight the potential of \planner for real-world applications such as warehouse automation and autonomous fleet coordination, where dynamic task assignment and collision-free movements are crucial. 
Ensuring collision-free movement under computational delays further enhances its robustness in practical scenarios.

In conclusion, this work contributes to the field of MAPF by offering a practical solution for the lifelong MAPF problem in continuous time with volumetric agents.